%% file: ANA-EXOT-2021-32-PAPER.tex
\documentclass[cernpreprint, atlasdraft=false, UKenglish, texmf, orcidlogo]{atlasdoc}
\usepackage{atlaspackage}
\usepackage{atlasbiblatex}

\usepackage{atlasphysics}
\graphicspath{{logos/}{figures/}}

\usepackage{ANA-EXOT-2021-32-PAPER-defs}

\AtlasTitle{Search for light long-lived particles in $pp$ collisions at
$\sqrt{s}=13$~TeV using displaced vertices in the ATLAS inner detector}

\AtlasVersion{2.0}

\AtlasAbstract{
A search for long-lived particles (LLPs) using 140~fb$^{-1}$ of $pp$ collision
data with $\sqrt{s} = 13$~TeV recorded by the ATLAS experiment at the LHC is
presented. The search targets LLPs with masses between $5$ and $55$~GeV
that decay hadronically in the ATLAS inner detector.
Benchmark models with LLP pair production from exotic decays of the Higgs boson
and models featuring long-lived axion-like particles (ALPs) are considered.
No significant excess above the expected background is observed.
Upper limits are placed on the branching ratio of the Higgs boson to pairs of
LLPs, the cross-section for ALPs produced in association with a vector boson,
and, for the first time, on the branching ratio of the top quark to an ALP and a
$u/c$ quark.}
\AtlasRefCode{EXOT-2021-32}

\PreprintIdNumber{CERN-EP-2024-086}

\AtlasJournal{Phys. Rev. Lett.}
\AtlasJournalRef{\PRL 133 (2024) 161803}
\AtlasDOI{10.1103/PhysRevLett.133.161803}

\hypersetup{pdftitle={ATLAS document},pdfauthor={The ATLAS Collaboration}}

\begin{document}

\maketitle

%
%
%

%

%
%
%
%
%
%

The majority of experimental searches for dark matter (DM) have concentrated on
weakly interacting massive particles (WIMPs) that interact directly with
Standard Model (SM) particles with a strength comparable to that of the weak interaction.
However, constraints on
WIMP-like DM from both
direct~\cite{PhysRevLett.119.181302,PhysRevD.100.022001,PhysRevD.98.102006,PhysRevD.100.102002,PhysRevLett.121.081307,PhysRevLett.118.021303,PhysRevLett.126.091301,Lai_2023,PhysRevD.99.062001}
and indirect detection experiments~\cite{Ackermann_2015}
are becoming increasingly stringent.  One compelling alternative to the WIMP
paradigm is that DM particles belong to a ``dark sector'' (DS) that is neutral
under the SM gauge group and interacts with the SM only via one or more beyond
the SM mediator
particles~\cite{HV_DV,EchoesHV,HVPhenoHadronColl,Schwaller_2015,Renner_2018}.
If decays of the mediator to DS particles are kinematically forbidden, its
decay back into SM particles will be suppressed by the small coupling
between the SM and the mediator, giving rise to
potentially macroscopic proper decay lengths ($c\tau \gtrsim 100$~$\mu$m).
These so-called long-lived particles (LLPs) are also predicted in scenarios in
which the mediator particle couples to the SM via a higher-dimensional operator,
such as in models featuring axion-like particles (ALPs)~\cite{Hochberg_2018,Bharucha_2023}.

This Letter presents a search for neutral LLPs that decay hadronically, giving a
distinct signature of one or more hadronic jets originating at a significantly
displaced position from the proton--proton ($pp$) collision point, referred to as
a displaced vertex (DV).
Three benchmark models are explored, motivated by different interactions between the SM and DS states.
The first benchmark considers the “Higgs Portal”, in which the
SM Higgs boson mediates interactions with the DS through its coupling to
a neutral spin-0 boson, $s$~\cite{Curtin_2014,Curtin_2015}.
This benchmark gives rise to exotic decays of the Higgs boson to a pair of
long-lived $s$ particles that decay back to SM particles with Yukawa-ordered branching ratios.
The search targets Higgs boson production either in association with a vector boson
($W/Z$) or via the vector boson fusion (VBF) process in which a quark from each
of the incoming protons radiates a heavy vector boson which then fuse to produce
a Higgs boson~\cite{deFlorian:2016spz}.
The second benchmark considers extending the SM with an ALP, $a$, which couples
to gluons and $W/Z$ bosons through effective dimension-5 operators, while
couplings to photons are suppressed.  These interactions are characterized by a
scale $f_a$ and Wilson coefficients $C_{\tilde{G}}$ and $C_{\tilde{W}}$,
respectively~\cite{Brivio_2017}.
These operators give rise to the production of $a$ in association with a vector boson
($W/Z$) and its subsequent decay exclusively into gluons.
The third benchmark considers an ALP, $a$, which couples to up-type quarks~\cite{Carmona_2021,Carmona_2022},
giving rise to exotic decays of the top quark $t\rightarrow ac/au$
in $t\overline{t}$ events.  In this model the $a$ boson decays predominantly into charm quark pairs
or gluons, with branching ratios that depend on $m_a$.
Example Feynman diagrams of the three benchmark processes can be found in Appendix A.

This search is performed with 140 fb$^{-1}$ of 13~\TeV $pp$
collision data collected by the ATLAS experiment at the Large Hadron Collider
(LHC)~\cite{Evans:2008zzb} from 2015 to 2018.  Several previous searches for Higgs boson decays to
LLPs have been performed that in combination exclude branching ratios $\mathrm{BR}(H \rightarrow ss) >
10\%$ for $s$ masses above 40~\GeV and proper decay lengths
between $10^{-3}$~m and
$10$~m~\cite{EXOT-2019-24,EXOT-2019-23,EXOT-2018-61,EXOT-2018-57,CMS-EXO-19-021,CMS-EXO-20-015,CMS-EXO-20-003,CMS-EXO-21-008}.
However, for masses below 40~\GeV, Higgs boson decays to LLPs with proper decay lengths below 100~mm
are unconstrained beyond the limit of 12\% on the branching ratio of the Higgs boson to undetected states~\cite{HIGG-2021-23}.
A limiting factor in probing this region of phase
space with the ATLAS experiment has been the reconstruction of displaced tracks
in the inner tracking detector (ID). In 2022, an improved version of the track
reconstruction pass for large-impact parameter tracks was deployed in
ATLAS~\cite{IDTR-2021-03}. This upgrade significantly reduced the rate of
reconstructing so-called fake tracks due to random hit combinations, thereby
enhancing computational efficiency and enabling the application of this reconstruction to every recorded data event.
This Letter reports the first direct application of this new track reconstruction, which significantly expands the
reach of this search with respect to previous ATLAS results and allows for sensitivity to previously unexplored phase
space.
Notably, this is the first search for Higgs boson decays to hadronically decaying LLPs in the ID to probe the VBF topology, and the first
search to probe hadronically decaying long-lived ALPs produced in association with a vector boson and via exotic decays of the
top quark.


%
%
%


%
\newcommand\AtlasCoordFootnote{ATLAS uses a right-handed coordinate system with its origin at the
nominal interaction point (IP) in the center of the detector and the $z$-axis
along the beam pipe. The $x$-axis points from the IP to the center of the LHC
ring, and the $y$-axis points upward. Polar coordinates $(r,\phi)$ are used in
the transverse plane, $\phi$ being the azimuthal angle around the $z$-axis. The
pseudorapidity is defined in terms of the polar angle $\theta$ as
$\eta=-\ln\tan(\theta/2)$. Angular distance is measured in units of $\Delta R = \sqrt{\Delta \eta^2 +\Delta \phi^2 }$.}

The ATLAS detector~\cite{PERF-2007-01,ATLAS-TDR-19} is a cylindrical detector with
forward--backward symmetry and nearly \(4\pi\) solid-angle
coverage.\footnote{\AtlasCoordFootnote}
It consists of the ID surrounded by a thin superconducting
solenoid, electromagnetic and hadronic calorimeters, and a muon spectrometer.
The ID covers the pseudorapidity range \(|\eta| < 2.5\) and consists of silicon
pixel, silicon microstrip, and transition radiation tracking detectors.
An extensive software
suite~\cite{ATL-SOFT-PUB-2021-001} is used in data simulation, in the
reconstruction and analysis of real and simulated data, in detector operations,
and in the trigger and data acquisition systems of the experiment.
A primary charged particle (track) reconstruction pass is used to
reconstruct charged-particle trajectories with transverse impact parameter
($d_0$) with respect to the $pp$ interaction point (IP) of $|d_0|<5$~mm.
A large-impact parameter pass, using left-over hits from the primary pass, is used to increase tracking acceptance up to
$|d_0|<300$~mm~\cite{IDTR-2021-03}.
The $pp$ interaction vertex with the highest sum of squared transverse momenta of associated tracks is taken as the primary interaction vertex
(PV).  Hadronic jets are reconstructed from topological clusters of energy deposits in the
calorimeters~\cite{JETM-2018-05} using the \antikt
algorithm~\cite{Cacciari:2008gp,Fastjet} with a radius parameter of $R = 0.4$. The
matching of tracks with the calorimeter-based jets is performed via the
ghost-association technique~\cite{Cacciari_2008}. Jets with transverse momentum $\pt > 20$~\GeV are
considered in the analysis.

Electron candidates are reconstructed from energy deposits in the calorimeters
associated to an ID track, and are required to be within the fiducial region
$|\eta| < 2.47$, and outside of $1.37 < |\eta| < 1.52$. Muons are reconstructed by combining tracks reconstructed in
the ID with tracks or track segments found in the muon spectrometer (MS) and are required to have
$|\eta| < 2.5$. Electrons and muons are required to have $\pt > 10$~\GeV and satisfy
the \emph{Medium} identification criterion~\cite{EGAM-2018-01,MUON-2018-03}. To
ensure that the selected electrons (muons) originate from the PV, they must
satisfy $|\frac{d_0}{\sigma(d_0)} | < 5 \, (3)$, and $|(z_0 - z_\textrm{PV})
\sin \theta| < 0.5$~mm, where $z_0$ is the track's longitudinal impact parameter
and $z_\textrm{PV}$ is the $z$ coordinate of the PV.
In this Letter, electrons and muons satisfying the above criteria will collectively be
referred to as \textit{leptons}.
Photon candidates are reconstructed from
clustered energy deposits in the electromagnetic calorimeter
either without any matching ID track or with a matching photon conversion vertex in the ID material.
The \emph{Loose} identification criterion is required~\cite{EGAM-2018-01}.
The missing transverse momentum (\met) is defined as the
magnitude of the negative vector sum of the transverse momenta of all
reconstructed and calibrated electrons, muons, photons, jets, and
remaining unclustered energy. The latter is estimated from low-\pt
tracks associated with the PV but not assigned to a reconstructed object~\cite{ATLAS-CONF-2018-023}.

Samples of Monte Carlo (MC) simulated events are used to study the three benchmark scenarios.
Signal samples were generated assuming mean proper decay lengths of the LLP
(either $s$ or $a$) of 1, 10, 100, and 1000~mm and masses of $m_s =$ 5, 16, 40,
55~\GeV and $m_a = $ 40, 55~\GeV for the Higgs portal and ALP benchmarks,
respectively.
In all samples, the LLP is taken to be a pseudoscalar,
although the analysis does not explicitly exploit the $CP$ properties of the
LLPs.
In the Higgs portal benchmark, the decays of the $s$ particles are simulated assuming a 100\% branching ratio to the
heaviest quark--antiquark pair that is kinematically allowed. To quantify the
dependence of the analysis on the flavor of the final state quarks, additional samples are generated
assuming a 100\% branching ratio to $u\overline{u}$.
In the $W/Z$ plus ALP benchmark, the decay of the $a$ particle is simulated assuming a 100\% branching ratio to $gg$.
In the exotic top decay benchmark, the $a$ particle can decay into either
$c\overline{c}$ or $gg$.
For the values of $m_a$ considered in this analysis, the branching ratios to $c\overline{c}$ and $gg$
are approximately 75\% and 25\%, respectively~\cite{Carmona_2022}.
Samples of simulated \ttbar and \(V+\)jets background events are used to
optimize the event selections and evaluate systematic uncertainties.
Details about the event simulation configurations used can be found in Appendix A.


%
%
%

%
Events are categorized into three search regions, each targeting a different
Higgs boson or ALP production mode. The \textit{1-lepton} region is defined by the presence of
exactly one lepton $(e/\mu)$ with $p_{T} > 27$~\GeV and $\met > 30$~\GeV.
These criteria target signal processes containing a leptonically decaying $W$ boson including $WH$, $Wa$, and $t\overline{t}, t \rightarrow ac/au$ production.
The \textit{2-lepton} region is defined by the presence of exactly two leptons, with the same flavor and opposite charge. The highest $\pt$ lepton is
required to have $p_{T} > 27$~\GeV, and the invariant mass of the dilepton
system is required to fall between 76 and 106~\GeV. These criteria target signal events containing a leptonically decaying $Z$ boson, including $ZH$ and $Za$ production.
Events in the 1- and 2-lepton regions are collected with a combination of single
and dilepton triggers~\cite{TRIG-2018-05,TRIG-2018-01,TRIG-2019-03}.
A matching requirement is applied between the selected leptons and the corresponding
leptons reconstructed by the trigger.
In both regions, events are required to have at least two jets with $|\eta| < 2.5$.
The \textit{VBF enriched} region targets the Higgs portal benchmark in the VBF production mode.
Events are collected with an inclusive VBF trigger~\cite{ATL-DAQ-PUB-2019-001} enabled during the 2018 data-taking period that
is designed to select events with a pair of jets consistent with the VBF
process. The data collected with this
trigger correspond to a total integrated luminosity of 37.5 fb$^{-1}$.
The VBF enriched region is defined by the absence
of any lepton, and the presence of a pair of jets with invariant mass
$m_{jj}>1200$~\GeV and angular separation $|\Delta\eta_{jj}| > 4$ and
$|\Delta\phi_{jj}| <2$. The leading (subleading) jet in this pair is required to
have transverse momentum $\pt> 100 \:(80)$~\GeV and $|\eta| < 3.2 \:(4.9)$. These
selections ensure that the trigger selection efficiency is
approximately 100\%.
In addition to the pair of jets used to select the VBF topology, events are required to have at least two additional jets with $|\eta| < 2.5$.

The jets emerging from the decay products of an LLP, referred to as \textit{displaced
jets}, exhibit a distinct topology compared to \textit{prompt jets} that originate from
a $pp$ interaction vertex. To distinguish displaced jets from prompt jets, a
per-jet boosted decision tree (BDT) is trained using the
\textsc{XGBoost} framework~\cite{Chen:2016:XST:2939672.2939785}.
The output of this classifier is a displaced jet BDT score between zero and one, where a higher score
indicates that the jet is more likely to have originated from a displaced decay.
This BDT is trained on five jet-level features that discriminate between displaced
and prompt jets.
The first feature is the fraction of the total jet \pt carried
by tracks with $|d_0| < 0.5$~mm, which is expected to be smaller for displaced
jets than for prompt jets. Similarly, the fraction of the total jet \pt carried
by tracks with $|d_0| > 0.5$~mm is used, which provides additional information
about the contribution from displaced charged particles to the total jet momentum.
Third, the fractional value of jet track \pt originating from tracks with $|d_0| < 0.5$~mm and
$|(z_0 - z_{\textrm{vertex}}) \sin{\theta}|<0.3$~mm is calculated for each reconstructed $pp$ interaction vertex,
and the maximum value of this set is taken.
Finally, the maximum $|d_0|$ among tracks in the jet, and the median of the logarithmic
transverse impact parameter significance of tracks associated to the jet are used.
The BDT is trained on a mixed signal sample comprised of $VH, H \rightarrow ss$ events with $m_s \in
\{16, 55\}$~\GeV and $c\tau_s \in \{10,100\}$~mm, and a mixed background sample
comprised of simulated \ttbar, \(W+\)jets and \(Z+\)jets events in equal parts.
Distributions of the BDT score for jets in selected signal samples and in data can be found in the Supplemental
Material.
To reconstruct the origin of the hadronic jets produced from the
decay of the LLPs, a DV reconstruction algorithm~\cite{ATL-PHYS-PUB-2019-013} is run on the
combined collection of tracks from both the primary and the large-impact
parameter tracking passes.
Following Ref.~\cite{EXOT-2018-57}, selections are placed on the reconstructed
vertices to reject DVs from SM processes and random combinations of tracks.
DVs are required to have a track multiplicity $n_{\text{track}} \geq 3$ and
vertex goodness of fit $\chi_{\text{DV}}^2/ n_{\text{DoF}} < 5$.
The radial and longitudinal coordinates of the DV position are each required to be less than
300~mm, and a material veto
is applied to reject DVs from interactions between high-momentum hadrons and
known detector elements~\cite{SUSY-2018-13}.
Furthermore, the minimum $|d_0|$ among all tracks associated to a DV ($|d_{0,\mathrm{min}}|$)
must satisfy $|d_{0,\mathrm{min}}|>0.1$~mm, and DVs must contain at least one
track with $|d_0|>3$~mm.

To compute the kinematic properties of DVs, the parameters of the tracks
associated to each DV are recalculated after extrapolating their
trajectories to the DV position. The resulting track four-momentum vectors,
measured with respect to the DV, are then summed together to yield the DV's
four-momentum. The ratio of the DV invariant mass ($m_\textrm{DV}$) and the maximum
angular distance between any two tracks in the DV ($\Delta R_{\text{max}}$) is
then required to satisfy $m_{\textrm{DV}}/\Delta R_{\text{max}} > 4$~\GeV, and
the scalar sum of the transverse momentum of tracks associated to a DV is
required to be above 10~\GeV. To associate DVs to displaced jets, the DV
momentum vector is required to be within $\Delta R < 0.6$ of a jet with a BDT
score greater than 0.5. If multiple DVs are matched to a given jet, only the DV
with the smallest $\Delta R$ to the jet axis is considered. DVs that satisfy all
of the selections above and are matched to a displaced jet are used to count the
DV multiplicity in the event ($n_{\mathrm{DV}}$).

All events considered in the analysis are required to have at least two jets
with a BDT score greater than 0.5.
An event-level discriminant ($\mathrm{BDT}_{j_0} \times \mathrm{BDT}_{j_1}$) is computed by
taking the product of the BDT scores of the two jets in the event with the
largest BDT scores.
From each of the three search regions, two signal regions (SRs) are defined
based on the candidate DV multiplicity in the event, $n_{\mathrm{DV}}=1$ or
$n_{\mathrm{DV}} \geq 2$, resulting in a total of six SRs.
Events in the
$n_{\mathrm{DV}} = 1$ SRs are required to have $\mathrm{BDT}_{j_0}
\times \mathrm{BDT}_{j_1} > 0.9$. This condition is relaxed to
$\mathrm{BDT}_{j_0} \times \mathrm{BDT}_{j_1} > 0.7$ in the $n_{\mathrm{DV}}
\geq 2$ SRs.
Example distributions of the event-level discriminant can be found in the Supplemental Material.

%


%
%
%

%
The dominant sources of background are $t\overline{t}$ and $W$+jets, $Z$+jets, and
multijet production in the 1-lepton, 2-lepton, and VBF enriched SRs,
respectively. The background contribution is estimated using a
fully data-driven approach, following the method developed in Ref.~\cite{EXOT-2018-57}. In each of the three search regions, a control region (CR)
is defined by requiring $\mathrm{BDT}_{j_0} \times \mathrm{BDT}_{j_1} < 0.7$.
Assuming a 12\% branching ratio of $H \rightarrow ss$ from Ref.~\cite{HIGG-2021-23}, the fractional signal contribution in the CRs
is expected to be less than 1\%. %
The probability that a jet is matched to a DV is computed separately in each of the three CRs and encoded in a
three dimensional map parameterized in jet \pt, the jet flavor tagging
score (DL1r)~\cite{FTAG-2019-07} that separates light and heavy flavor jets, and BDT score.
The map is divided evenly in the BDT
dimension using a bin width of 0.01 in the 1-lepton region, and 0.025 in the
2-lepton and VBF enriched regions where fewer events are selected.
The three probability maps are shown in Fig. 4 of the Supplemental Material as two-dimensional projections.
The per-jet probabilities are then used to
compute the probability that each event contains exactly one, or greater than
one DV based on the \pt, DL1r, and BDT scores of the jets in the event.
The per-event probability weights are applied inclusively to data in the
search regions to predict the distributions of $\mathrm{BDT}_{j_0} \times \mathrm{BDT}_{j_1}$ in events with $n_{\mathrm{DV}} =
1$ and $n_{\mathrm{DV}} \geq 2$.

Two uncertainties in the background prediction are considered. First, the
statistical uncertainty in the background estimate due to the finite number of events in the CR
used to derive the maps is computed using ensembles of background
estimates from a set of statistically varied per-jet probability
maps~\cite{EXOT-2018-57}. The standard deviation of this ensemble of estimates
is used to define the up and down statistical variations on the nominal
prediction.
Second, in the 2-lepton and VBF enriched regions, where a coarser binning is
used in the BDT dimension of the per-jet probability map, an uncertainty in the
background estimate from the binning choice is quantified.
In this regard,
the difference between the nominal estimate and an alternate estimate computed from a map with a BDT bin width of 0.01
is taken as a systematic uncertainty.
The total uncertainty in
the background predictions varies from 10--50\%, depending on the signal region.
The largest uncertainties are present in the $n_{\textrm{DV}} \geq 2$ regions, especially in the 2-lepton and VBF
regions, where the statistical uncertainty is dominant due to the finite number of jets available for deriving the per-jet probability maps.

The background estimate is validated in a subset of the $n_{\mathrm{DV}} = 1$
events defined by $0.7 < \mathrm{BDT}_{j_0} \times \mathrm{BDT}_{j_1} < 0.9$
within each of the three search regions, and in a dedicated event selection
requiring the presence of a single photon with $\pt> 160$~\GeV and $|\eta|<
2.47$, zero leptons, and two jets with $\pt > 20$~\GeV. The distributions of data events are found to be well modeled by the
predicted background in all regions, validating the extrapolation of the
background estimate from the CR to larger values of $\mathrm{BDT}_{j_0} \times
\mathrm{BDT}_{j_1}$ and to events with $n_{\mathrm{DV}} \geq 2$.  A more
detailed description of the background validation is given in Appendix B.


%
%
%

Instrumental and theoretical uncertainties are assigned on the modeling of the
simulated signal samples.
The dominant systematic uncertainty is due to the modeling of the BDT score, which is derived as a per-jet uncertainty by comparing
the shape of the BDT score between data and the
simulated $Z$+jets sample in the 2-lepton preselection, and then propagated to the final event yield.
The impact of this uncertainty is approximately 15\%.
In the VBF selection, the dominant systematic uncertainty is on the jet energy scale and resolution, reaching values of up to 20\%
due to the increased uncertainty associated with calibrating jets which have large pseudorapidity~\cite{JETM-2018-05}.
Subleading sources of instrumental uncertainty
include those on the primary and large-impact parameter track reconstruction efficiencies (2-9\%)~\cite{IDTR-2021-03}; lepton trigger,
reconstruction, and identification efficiencies (0-2\%)~\cite{EGAM-2021-01}; lepton energy scale and
resolution (0-1\%)~\cite{EGAM-2021-02};  modeling of the pileup in
simulation (2-4\%)~\cite{STDM-2015-05}; and the total integrated luminosity of the
measurement (0.8\%)~\cite{DAPR-2021-01,LUCID2}.
Theoretical uncertainties are considered to account for variations due to
the renormalization and factorization scales, parton distribution functions, and
parton showering (2-8\%).

\begin{figure}[ht!]
\centering
\includegraphics[width=0.8\textwidth]{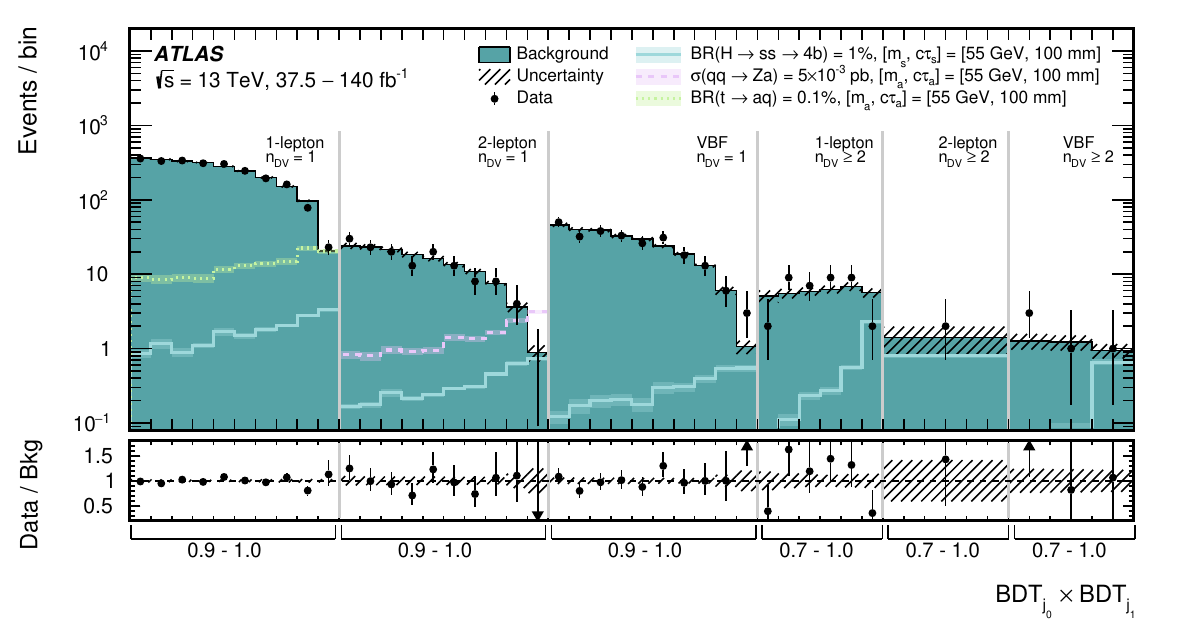}
\caption{Distributions of $\mathrm{BDT}_{j_0} \times \mathrm{BDT}_{j_1}$ for the observed data (black points) and the
estimated background (filled histogram) with its uncertainty after the background-only fit to data in the six SRs described in the text.
The signal expectation for the Higgs portal model with $m_{s}=55$~\GeV and $c\tau_{s}=100$~mm
is shown in the solid line, scaled to $\textrm{BR}(H\rightarrow ss \rightarrow 4b)=1\%$.
The signal expectation for ALP production in association with a $Z$ boson scaled to $\sigma(qq\rightarrow Za)=5\times
10^{-3}$~pb, and for $t \rightarrow aq$ scaled to $\mathrm{BR}(t \rightarrow aq) = 0.1\%$ are shown in the dashed and dotted lines,
respectively, for $m_{a}=55$~\GeV and $c\tau_{a}=100$~mm.
The observed data in the 1- and 2-lepton (VBF) regions corresponds to an integrated luminosity of 140 (37.5) fb$^{-1}$.
The ratio between the data and estimated background is shown in the bottom panel.
}
\label{fig:SRPlot}
\end{figure}

For each signal model considered, a binned maximum-likelihood fit to the
$\mathrm{BDT}_{j_0} \times \mathrm{BDT}_{j_1}$ distributions in the SRs is
performed. For the Higgs portal model, all six SRs are fitted simultaneously,
while in the $Wa$ and $t \rightarrow aq$ ($Za$) model, only the 1-lepton (2-lepton)
$n_{\mathrm{DV}} = 1$ SR is considered.
Systematic and MC statistical uncertainties are included as nuisance
parameters and are constrained in the fit. Systematic uncertainties on the
signal efficiency are correlated across the six signal regions, while the
systematic and statistical uncertainties on the background are treated as uncorrelated.  In the 1-lepton region where
there is no systematic uncertainty on the background prediction, a shape uncertainty is included as an additional degree
of freedom to the fit. This uncertainty varies linearly across the $n_{\mathrm{DV}}=1$ and $n_{\mathrm{DV}}\geq2$ SRs
and allows for the value of the fitted statistical uncertainty to vary from bin-to-bin.

\begin{figure}[ht!]
\centering
\includegraphics[width=0.35\textwidth]{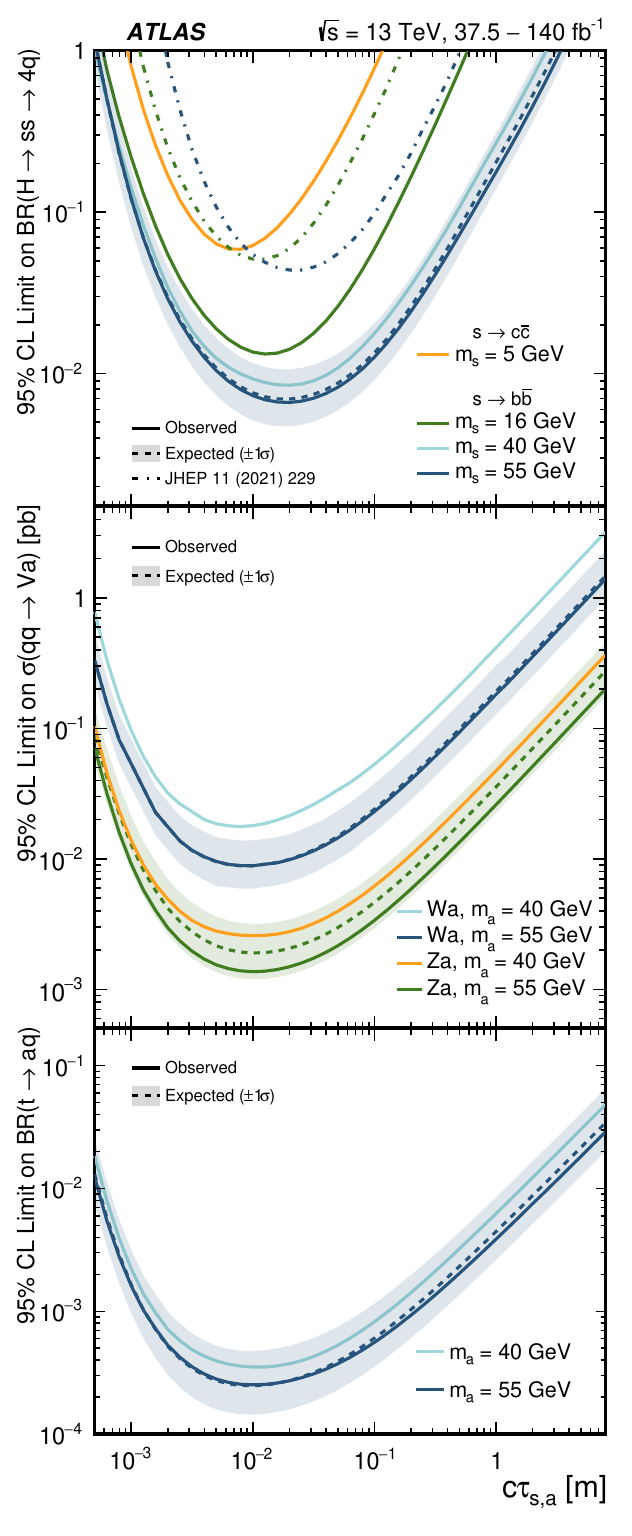}
\caption{
The 95\% confidence level limits on the (upper) Higgs boson branching ratio $H \rightarrow ss \rightarrow 4q$, (middle)
$q\overline{q}\rightarrow Va$ cross-section where $V=W$~or~$Z$, and (lower) $t
\rightarrow aq$ branching ratio shown as a function of the mean proper decay length $c\tau$ of the long-lived
particle.
The observed limits are shown
with a solid line. The expected limits and corresponding $\pm1\sigma$ uncertainty bands
for $m_{s/a} = 55$~GeV are shown with dashed lines and shaded bands, respectively.
In the upper plot, the limits shown are on the Higgs boson branching
ratio $H \rightarrow ss \rightarrow 4b$ for $m_s=16,40,55$~\GeV, and $H
\rightarrow ss \rightarrow 4c$ for $m_s=5$~\GeV.
In the upper plot, the observed limits for the Higgs Portal
model from the previous ATLAS search~\cite{EXOT-2018-57} are shown
with the dot-dashed lines.}
\label{fig:limits}
\end{figure}

The distributions of $\mathrm{BDT}_{j_0} \times
\mathrm{BDT}_{j_1}$ for the observed data and the background prediction after the background-only fit to data in the six SRs are shown in
Figure~\ref{fig:SRPlot}. No significant deviation from the SM expectation is
observed. The absence of a data excess is translated into exclusion limits at
95\% confidence level (CL) on $\textrm{BR}(H \rightarrow ss \rightarrow 4b)$, $\sigma(qq
\rightarrow Va)$, and $\textrm{BR}(t \rightarrow aq)$. The CLs
prescription~\cite{Read:2002hq} is used to compute the limits using asymptotic
formulae for the profile likelihood ratio~\cite{Cowan:2010js}.
The signal yields at different lifetimes are computed by reweighting the
exponential LLP decay distributions from the generated values to each target
lifetime, following the procedure described in Ref.~\cite{EXOT-2018-57}.
The expected and observed exclusion limits are shown in Figure~\ref{fig:limits}.
A comparison of the observed exclusion limits for light and heavy-flavor quark final states can be found in the
Supplemental Material for $H \rightarrow ss \rightarrow 4q$, as well as an interpretation of the limits on $\sigma(qq
\rightarrow Va)$ in terms of the Wilson coefficients $C_{\tilde{W}}$ and $C_{\tilde{G}}$ parameterizing the effective
$aVV$ and $agg$ vertices~\cite{Brivio_2017}.
For the Higgs portal benchmark, the limits set by this search are considerably
stronger than previous ATLAS results~\cite{EXOT-2018-57} using the same dataset, with improvements by
as much as a factor of 20 for $m_s = 55$~\GeV and $c\tau_s < 10$~mm. These
improvements are driven by the updated large-impact parameter track
reconstruction, the addition of the 1-lepton and VBF search regions, and the inclusion of $n_{\mathrm{DV}} = 1$ signal regions.
In particular, the VBF enriched region contributes a similar level of sensitivity as the other two search regions
despite having a considerably smaller integrated luminosity.


%
%
%
\FloatBarrier

%
%
%
%
%
%
%
%
%
%
%
%

%
%
%
%
%
%
%
%
%
%
%

In summary, this Letter reports the results of a search for LLPs with masses
between $5$ and $55$~\GeV that decay hadronically in the ATLAS inner detector.  No
significant excess beyond the SM prediction is observed.
The reported constraints on the Higgs boson branching ratio are the most
stringent to date for $m_s < 40$~\GeV and $1 < c\tau_s < 100$~mm.
For $H \rightarrow ss
\rightarrow 4b$, branching ratios greater than 1\% are excluded for $m_s =
55$~\GeV and $5.4 < c\tau_s < 72$~mm.  The exclusion limits are stronger for
light-quark final states, with $H\rightarrow ss \rightarrow 4u$ branching ratios
greater than 1\% excluded for $m_s =
55$~\GeV and $4.2 < c\tau_s < 110$~mm.  For the first time at the
LHC, branching ratios beyond the limit of 12\% imposed on Higgs boson decays to
undetected states are probed for $m_s < 16$~\GeV and $c\tau_s < 100$~mm, with
$\mathrm{BR}(H \rightarrow ss \rightarrow 4c) > 10\%$ excluded for $m_s=5$~\GeV
and $2.9 < c\tau_s < 21$~mm.  The first limits on long-lived ALP models with
suppressed coupling to photons are set, excluding cross-sections for
$qq\rightarrow Va$ greater than 0.1 pb for $40 < m_a < 55$~\GeV and $1.0 < c\tau_a < 220$~mm.
Long-lived ALPs produced via $t \rightarrow aq$  are probed for
the first time, excluding $t \rightarrow aq$ branching ratios greater than 0.1\%
between $1.6 < c\tau_a < 130$~mm for $40 < m_a < 55$~\GeV.

\FloatBarrier
\section*{Acknowledgments}
\FloatBarrier

%
%

%
%

We thank CERN for the very successful operation of the LHC and its injectors, as well as the support staff at
CERN and at our institutions worldwide without whom ATLAS could not be operated efficiently.

The crucial computing support from all WLCG partners is acknowledged gratefully, in particular from CERN, the ATLAS Tier-1 facilities at TRIUMF/SFU (Canada), NDGF (Denmark, Norway, Sweden), CC-IN2P3 (France), KIT/GridKA (Germany), INFN-CNAF (Italy), NL-T1 (Netherlands), PIC (Spain), RAL (UK) and BNL (USA), the Tier-2 facilities worldwide and large non-WLCG resource providers. Major contributors of computing resources are listed in Ref.~\cite{ATL-SOFT-PUB-2023-001}.

We gratefully acknowledge the support of ANPCyT, Argentina; YerPhI, Armenia; ARC, Australia; BMWFW and FWF, Austria; ANAS, Azerbaijan; CNPq and FAPESP, Brazil; NSERC, NRC and CFI, Canada; CERN; ANID, Chile; CAS, MOST and NSFC, China; Minciencias, Colombia; MEYS CR, Czech Republic; DNRF and DNSRC, Denmark; IN2P3-CNRS and CEA-DRF/IRFU, France; SRNSFG, Georgia; BMBF, HGF and MPG, Germany; GSRI, Greece; RGC and Hong Kong SAR, China; ISF and Benoziyo Center, Israel; INFN, Italy; MEXT and JSPS, Japan; CNRST, Morocco; NWO, Netherlands; RCN, Norway; MNiSW, Poland; FCT, Portugal; MNE/IFA, Romania; MESTD, Serbia; MSSR, Slovakia; ARRS and MIZ\v{S}, Slovenia; DSI/NRF, South Africa; MICINN, Spain; SRC and Wallenberg Foundation, Sweden; SERI, SNSF and Cantons of Bern and Geneva, Switzerland; MOST, Taipei; TENMAK, T\"urkiye; STFC, United Kingdom; DOE and NSF, United States of America.

Individual groups and members have received support from BCKDF, CANARIE, CRC and DRAC, Canada; CERN-CZ, PRIMUS 21/SCI/017, FORTE and PRIMUS, Czech Republic; COST, ERC, ERDF, Horizon 2020, ICSC-NextGenerationEU and Marie Sk{\l}odowska-Curie Actions, European Union; Investissements d'Avenir Labex, Investissements d'Avenir Idex and ANR, France; DFG and AvH Foundation, Germany; Herakleitos, Thales and Aristeia programmes co-financed by EU-ESF and the Greek NSRF, Greece; BSF-NSF and MINERVA, Israel; Norwegian Financial Mechanism 2014-2021, Norway; NCN and NAWA, Poland; La Caixa Banking Foundation, CERCA Programme Generalitat de Catalunya and PROMETEO and GenT Programmes Generalitat Valenciana, Spain; G\"{o}ran Gustafssons Stiftelse, Sweden; The Royal Society and Leverhulme Trust, United Kingdom.

In addition, individual members wish to acknowledge support from CERN: European Organization for Nuclear Research (CERN PJAS); Chile: Agencia Nacional de Investigaci\'on y Desarrollo (FONDECYT 1190886, FONDECYT 1210400, FONDECYT 1230812, FONDECYT 1230987); China: Chinese Ministry of Science and Technology (MOST-2023YFA1605700), National Natural Science Foundation of China (NSFC - 12175119, NSFC 12275265, NSFC-12075060); Czech Republic: Czech Science Foundation (GACR - 24-11373S), Ministry of Education Youth and Sports (FORTE CZ.02.01.01/00/22\_008/0004632), PRIMUS Research Programme (PRIMUS/21/SCI/017); EU: H2020 European Research Council (ERC - 101002463); European Union: European Research Council (ERC - 948254, ERC 101089007), Horizon 2020 Framework Programme (MUCCA - CHIST-ERA-19-XAI-00), European Union, Future Artificial Intelligence Research (FAIR-NextGenerationEU PE00000013), Italian Center for High Performance Computing, Big Data and Quantum Computing (ICSC, NextGenerationEU); France: Agence Nationale de la Recherche (ANR-20-CE31-0013, ANR-21-CE31-0013, ANR-21-CE31-0022), Investissements d'Avenir Labex (ANR-11-LABX-0012); Germany: Baden-Württemberg Stiftung (BW Stiftung-Postdoc Eliteprogramme), Deutsche Forschungsgemeinschaft (DFG - 469666862, DFG - CR 312/5-2); Italy: Istituto Nazionale di Fisica Nucleare (ICSC, NextGenerationEU); Japan: Japan Society for the Promotion of Science (JSPS KAKENHI JP21H05085, JSPS KAKENHI JP22H01227, JSPS KAKENHI JP22H04944, JSPS KAKENHI JP22KK0227); Netherlands: Netherlands Organisation for Scientific Research (NWO Veni 2020 - VI.Veni.202.179); Norway: Research Council of Norway (RCN-314472); Poland: Polish National Agency for Academic Exchange (PPN/PPO/2020/1/00002/U/00001), Polish National Science Centre (NCN 2021/42/E/ST2/00350, NCN OPUS nr 2022/47/B/ST2/03059, NCN UMO-2019/34/E/ST2/00393, UMO-2020/37/B/ST2/01043, UMO-2021/40/C/ST2/00187, UMO-2022/47/O/ST2/00148, UMO-2023/49/B/ST2/04085); Slovenia: Slovenian Research Agency (ARIS grant J1-3010); Spain: Generalitat Valenciana (Artemisa, FEDER, IDIFEDER/2018/048), Ministry of Science and Innovation (MCIN \& NextGenEU PCI2022-135018-2, MICIN \& FEDER PID2021-125273NB, RYC2019-028510-I, RYC2020-030254-I, RYC2021-031273-I, RYC2022-038164-I), PROMETEO and GenT Programmes Generalitat Valenciana (CIDEGENT/2019/023, CIDEGENT/2019/027); Sweden: Swedish Research Council (Swedish Research Council 2023-04654, VR 2018-00482, VR 2022-03845, VR 2022-04683, VR 2023-03403, VR grant 2021-03651), Knut and Alice Wallenberg Foundation (KAW 2018.0157, KAW 2018.0458, KAW 2019.0447, KAW 2022.0358); Switzerland: Swiss National Science Foundation (SNSF - PCEFP2\_194658); United Kingdom: Leverhulme Trust (Leverhulme Trust RPG-2020-004), Royal Society (NIF-R1-231091); United States of America: U.S. Department of Energy (ECA DE-AC02-76SF00515), Neubauer Family Foundation.

%
%


%

%
%
\appendix
%

%
%

\section*{APPENDIX A: SIMULATED DATA SAMPLES}

\begin{figure}[!ht]
\begin{center}

\subfloat[]{\includegraphics[width=0.4\textwidth]{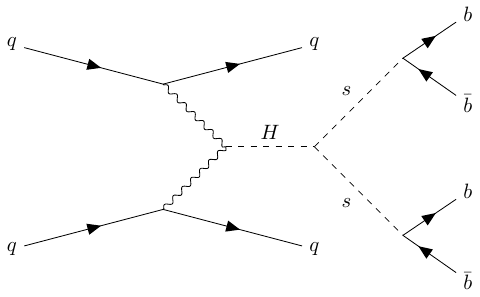}} \hspace{2cm}
\subfloat[]{\includegraphics[width=0.4\textwidth]{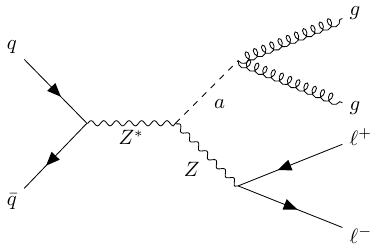}} \\
\subfloat[]{\includegraphics[width=0.4\textwidth]{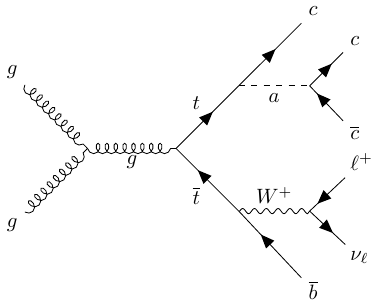}}

\caption{Example Feynman diagrams for the three benchmark models considered in
the analysis. Figure (a) shows an example diagram for the Higgs portal model, in
which the Higgs boson is produced via Vector Boson Fusion, and the long-lived
$s$ particles decay to pairs of $b$-quarks. Figure (b) shows an example of the $Va$
ALP production mode, in which the ALP $a$ is produced in association with a $Z$ boson,
with $a \rightarrow gg$ and $Z \rightarrow \ell^+ \ell^-$. Figure (c) shows an
example diagram of ALP production via the exotic top-quark decay, with $t \rightarrow ac$
and $a \rightarrow c\overline{c}$.}
\label{fig:feynman}
\end{center}
\end{figure}

Samples of MC simulated events are used to study the three benchmark scenarios. Higgs boson
production in association with a vector boson and via vector-boson fusion (VBF) was
simulated using
\POWHEGBOX[v2]~\cite{Frixione:2007nw,Nason:2004rx,Frixione:2007vw,Alioli:2010xd}
and interfaced with \PYTHIA[8.2]~\cite{Sjostrand:2014zea} to simulate the $H\rightarrow ss$ decay and the subsequent decay of
the $s$. \PYTHIA[8.2] is also used for simulating parton shower
and non-perturbative effects, with parameters set according to the \AZNLO
tune~\cite{STDM-2012-23}.  The \POWHEGBOX prediction is accurate to
next-to-leading-order (NLO) for \(VH\) boson plus one-jet production.  The
loop-induced \(gg\to ZH\) process was generated separately at leading-order (LO).
Samples of $pp \rightarrow Va$ and $pp \rightarrow t\overline{t}$ with $t \rightarrow ac/au$
were simulated using \textsc{MadGraph}~\cite{Alwall:2014hca} v2.9.9 and
interfaced with \textsc{Pythia} 8.307.
The effect of multiple $pp$ interactions in the same or neighboring
bunches (pileup) was modeled by overlaying the hard-scatter process with
simulated inelastic $pp$ scattering events.
Example Feynman diagrams of the simulated signal processes are shown in Figure~\ref{fig:feynman}.

The production of \ttbar events was modeled using the \POWHEGBOX[v2] generator at
NLO and interfaced with \PYTHIA[8.230] with parameters set according to the A14
tune~\cite{ATL-PHYS-PUB-2014-021}.  The production of \(V+\)jets was
simulated with the \SHERPA[2.2.1]~\cite{Bothmann:2019yzt} generator using NLO
matrix elements for up to two partons, and LO matrix elements for up
to four partons calculated with the Comix~\cite{Gleisberg:2008fv} and
\OPENLOOPS~\cite{Buccioni:2019sur,Cascioli:2011va,Denner:2016kdg} libraries.
They were matched with the \SHERPA parton shower~\cite{Schumann:2007mg} using
the \MEPSatNLO
prescription~\cite{Hoeche:2011fd,Hoeche:2012yf,Catani:2001cc,Hoeche:2009rj}. The
samples were normalized to a next-to-next-to-leading-order
prediction~\cite{Anastasiou:2003ds}.

\section*{APPENDIX B: VALIDATION OF BACKGROUND ESTIMATE}

The method of estimating the distributions of events with $n_{\mathrm{DV}}=1$ from the per-jet
probabilities is validated by performing closure tests in the CRs.
The weighted distributions of $\mathrm{BDT}_{j_0} \times \mathrm{BDT}_{j_1}$ in the
three CRs are found to reproduce the observed distributions of events with $n_{\mathrm{DV}}=1$
within uncertainties. The distributions of the leading and subleading jet $\pt$, DL1r, and BDT
scores are also found to be well modeled by the estimate in CR events with $n_{\mathrm{DV}}=1$.

\begin{figure}[h]
\centering
\includegraphics[width=0.5\textwidth]{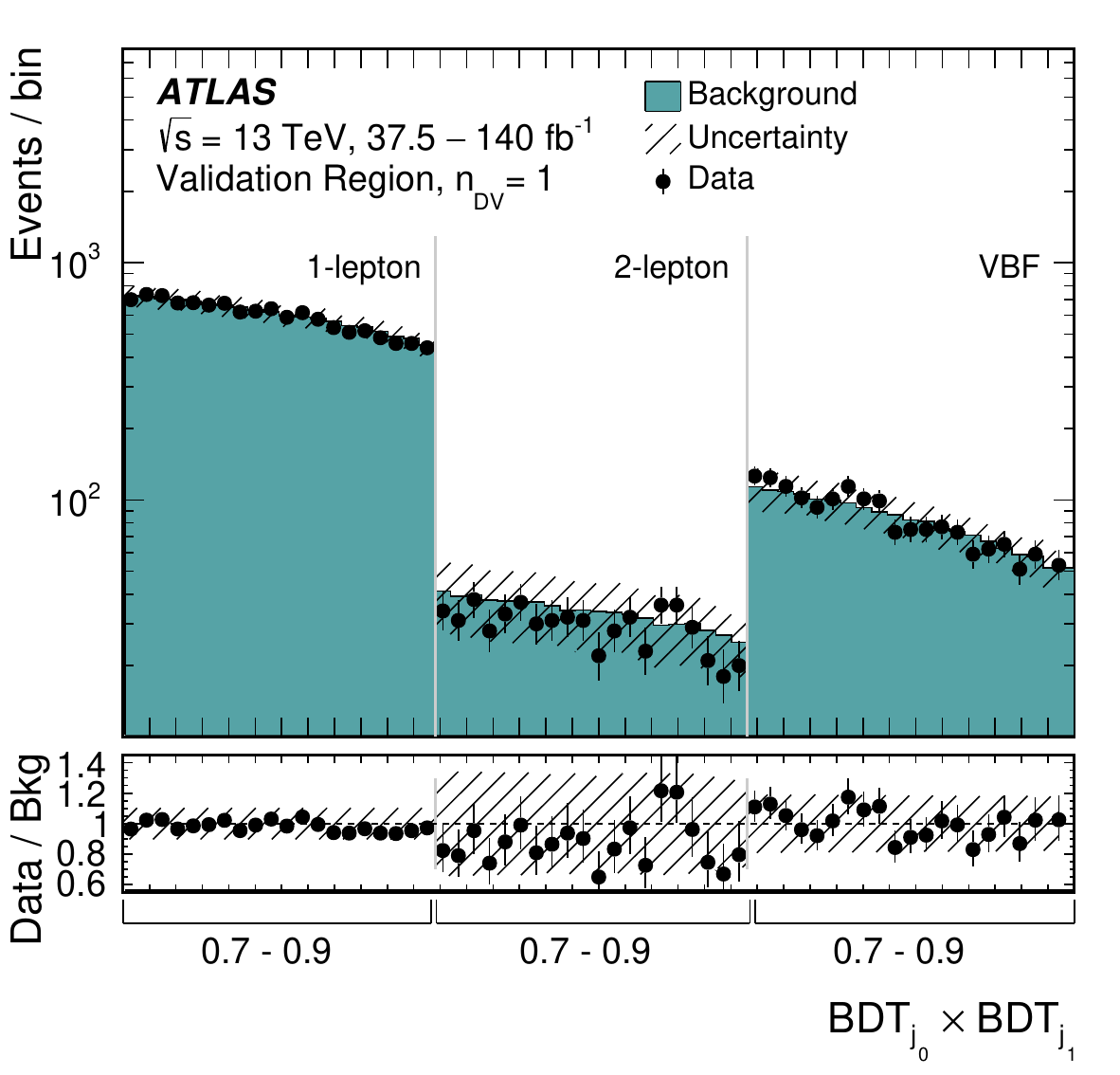}
\caption{Distributions of $\mathrm{BDT}_{j_0} \times \mathrm{BDT}_{j_1}$ for
the observed data (black points) and the background prediction (teal filled histogram) with its
uncertainty in the three $n_{\mathrm{DV}} = 1$ validation regions with
$0.7 < \mathrm{BDT}_{j_0} \times \mathrm{BDT}_{j_1} < 0.9$.
The ratio between the data and predicted background is shown in the bottom panel.
The background estimates are computed using
events in the three CRs with $\mathrm{BDT}_{j_0} \times \mathrm{BDT}_{j_1} < 0.7$.}
\label{fig:VRPlot}
\end{figure}

The extrapolation of the background estimate from the CR to larger values of
$\mathrm{BDT}_{j_0} \times \mathrm{BDT}_{j_1}$ is validated in a subset of the
$n_{\mathrm{DV}} = 1$ events defined by
$0.7 < \mathrm{BDT}_{j_0} \times \mathrm{BDT}_{j_1} < 0.9$, within the 1-lepton, 2-lepton and VBF enriched search regions.
The observed data in these three validation regions (VRs) are found to agree with the predicted
background within uncertainties, as shown in Figure~\ref{fig:VRPlot}.
The largest discrepancy is observed in the 2-lepton VR, with 590 events observed
compared to a predicted yield of $676 \pm 193 (\textrm{stat.}) \pm 106 (\textrm{syst.})$.

\begin{figure}[h]
\centering
\includegraphics[width=0.5\textwidth]{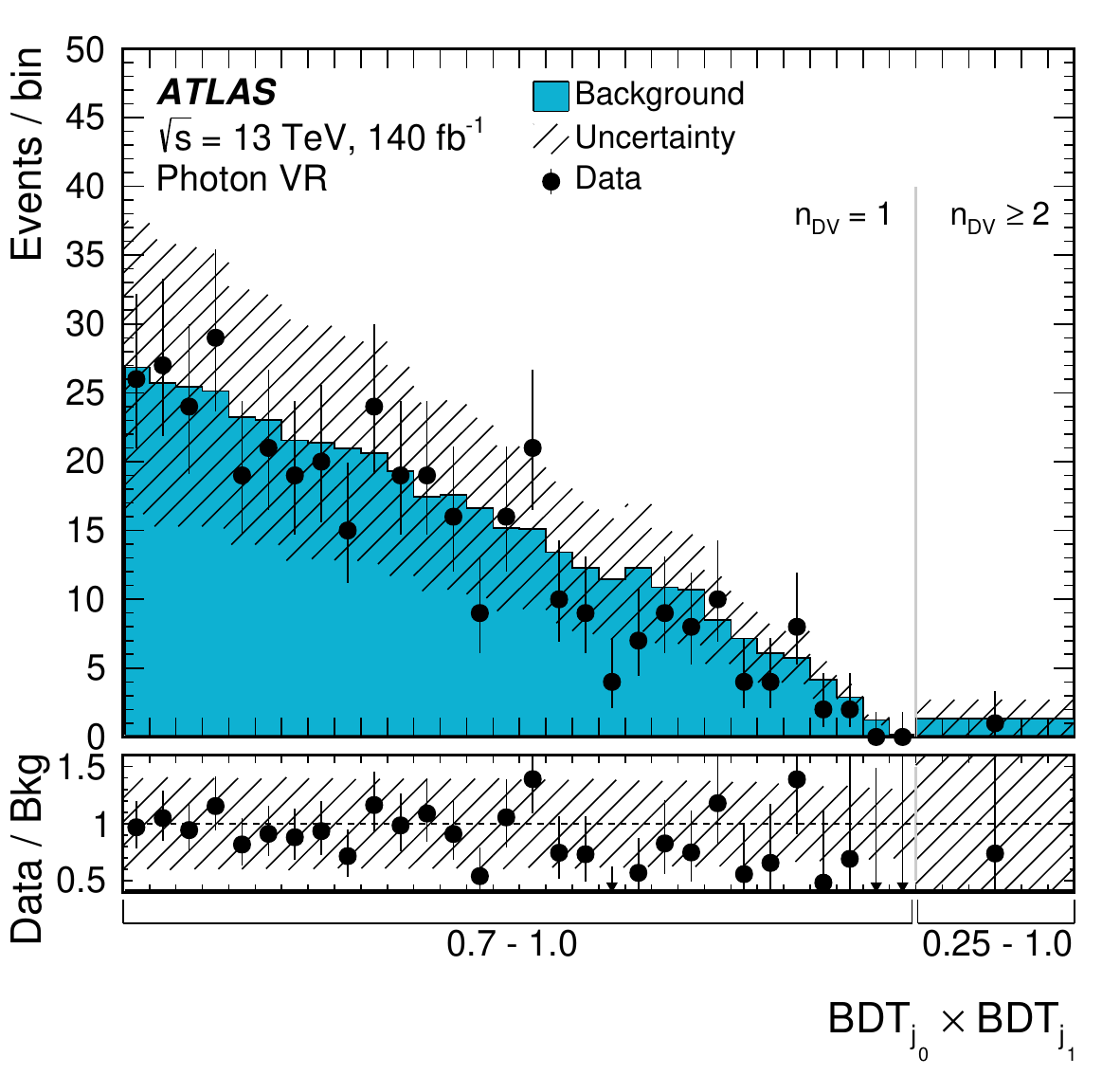}
\caption{Distributions of $\mathrm{BDT}_{j_0} \times \mathrm{BDT}_{j_1}$ for
the observed data (black points) and the background prediction (blue) with its
uncertainty in the photon validation region (VR) for events with $n_{\mathrm{DV}} =
1$ and $n_{\mathrm{DV}} \geq 2$.
The ratio between the data and predicted background is shown in the bottom panel.
The background estimate is computed using
events in the photon VR with $\mathrm{BDT}_{j_0} \times \mathrm{BDT}_{j_1} < 0.7$.}
\label{fig:VRPhot}
\end{figure}

To validate the extrapolation to higher values
of $\mathrm{BDT}_{j_0} \times \mathrm{BDT}_{j_1}$ and to events with
$n_{\mathrm{DV}} \geq 2$, a dedicated \textit{photon VR} is used defined by the presence of a single photon
with $\pt> 160$~\GeV and $|\eta|< 2.47$, zero leptons, and two jets with $\pt > 20$~\GeV. This selection effectively rejects events from all signal models considered
and provides an independent set of data events on which to test the background
estimation method.  The same background estimation strategy is applied to this
region as in the three search regions, using a dedicated map computed from
events in the photon VR with $\mathrm{BDT}_{j_0} \times \mathrm{BDT}_{j_1} <
0.7$.  The distributions of data events with $n_{\mathrm{DV}} = 1$ and
$\mathrm{BDT}_{j_0} \times \mathrm{BDT}_{j_1} > 0.7$ and data events with
$n_{\mathrm{DV}} \geq 2$ are found to be well modeled by the predicted
background, as shown in Figure~\ref{fig:VRPhot}.


%

%
%

%
%
%
\FloatBarrier
\printbibliography

\FloatBarrier
\clearpage

\clearpage
\part*{Supplemental Material}
\begin{figure}[!h]
\begin{center}
\begin{tabular}{cc}

\subfloat[]{\includegraphics[width=0.5\textwidth]{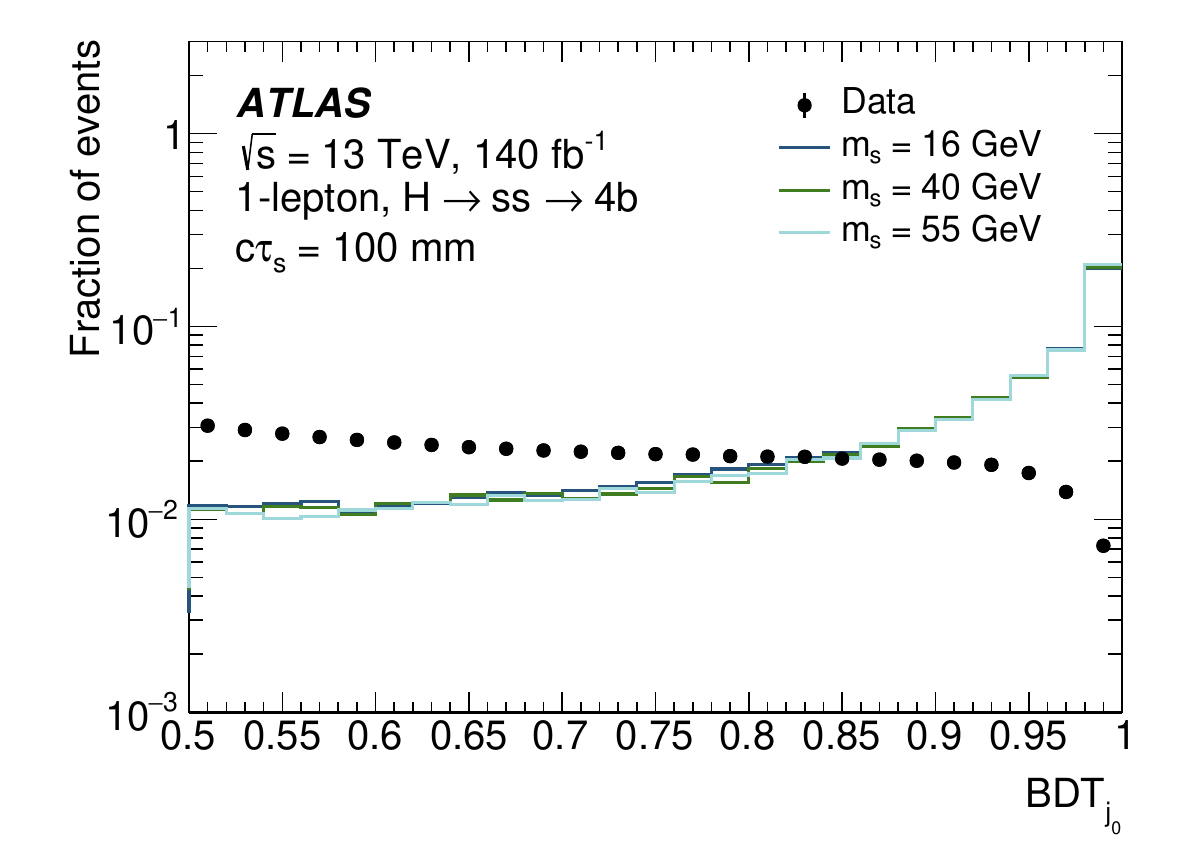}}
\subfloat[]{\includegraphics[width=0.5\textwidth]{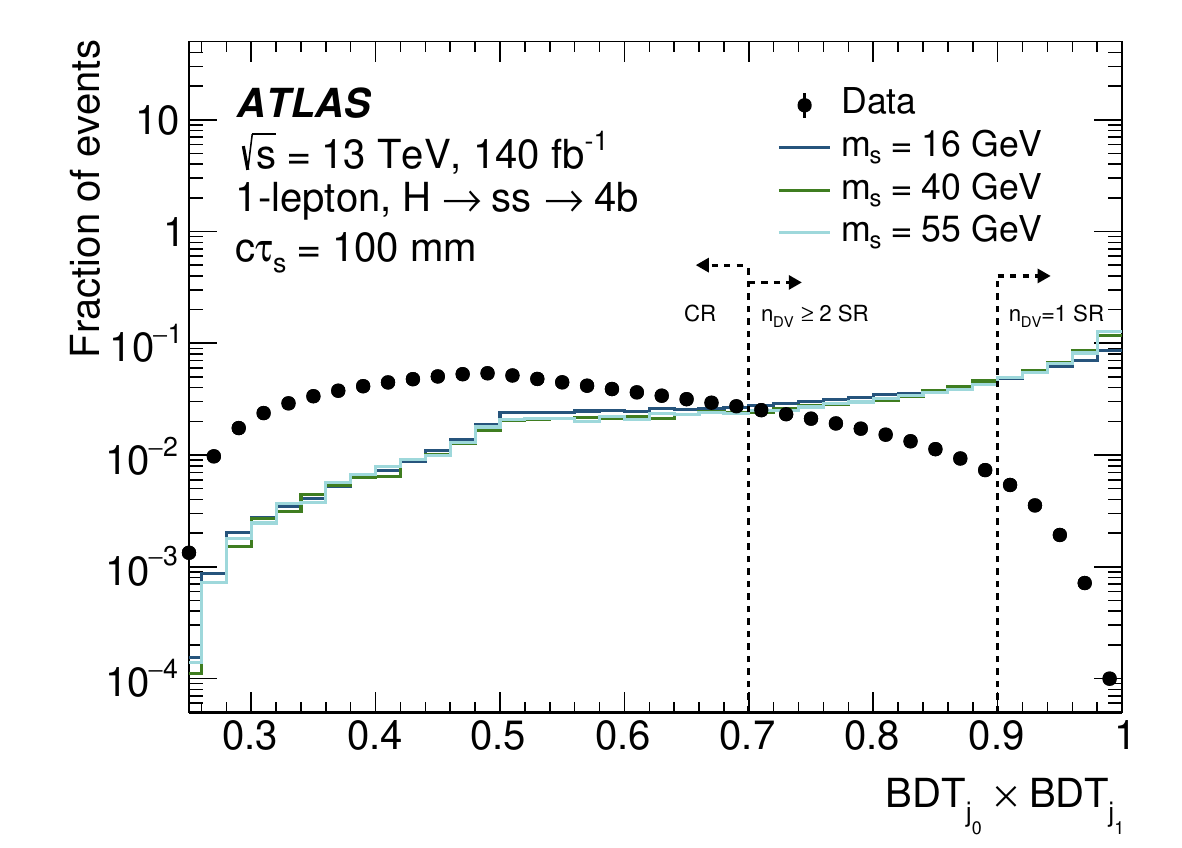}}

\end{tabular}
\caption{The distribution of (a) the BDT score of the jet with the highest BDT score in the event ($\mathrm{BDT}_{j_0}$) and
(b) the product of the BDT scores of the two
jets in the event with the largest BDT scores ($\mathrm{BDT}_{j_0} \times
\mathrm{BDT}_{j_1}$), in simulated Higgs portal signal samples with $H \rightarrow ss \rightarrow 4b$ (solid lines) and in data
(black points) for events in the 1-lepton search region described in the text,
with the additional requirement that the events contain at least two jets with
BDT scores greater than 0.5. The distributions are normalized to unit area.
}
\end{center}
\end{figure}

\begin{figure}[!h]
\begin{center}
\begin{tabular}{cc}

\subfloat[]{\includegraphics[width=0.5\textwidth]{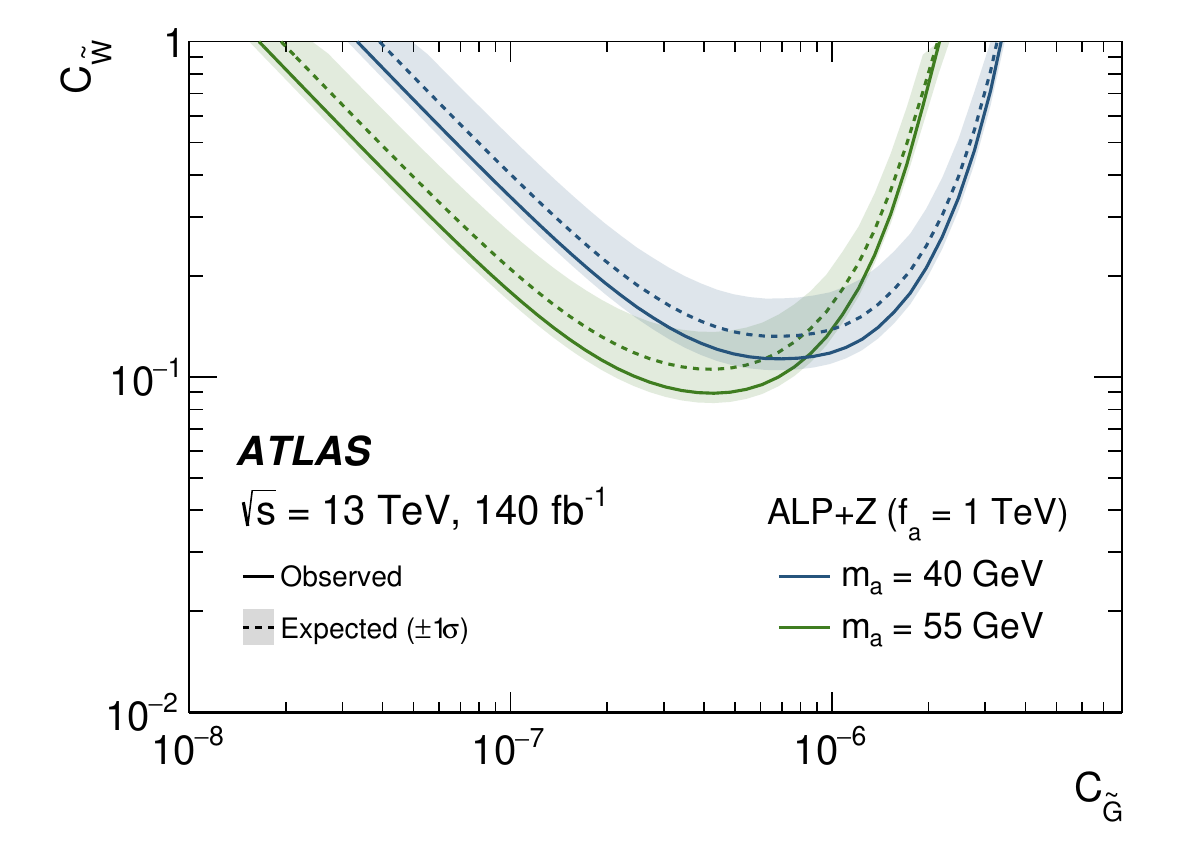}}
\subfloat[]{\includegraphics[width=0.5\textwidth]{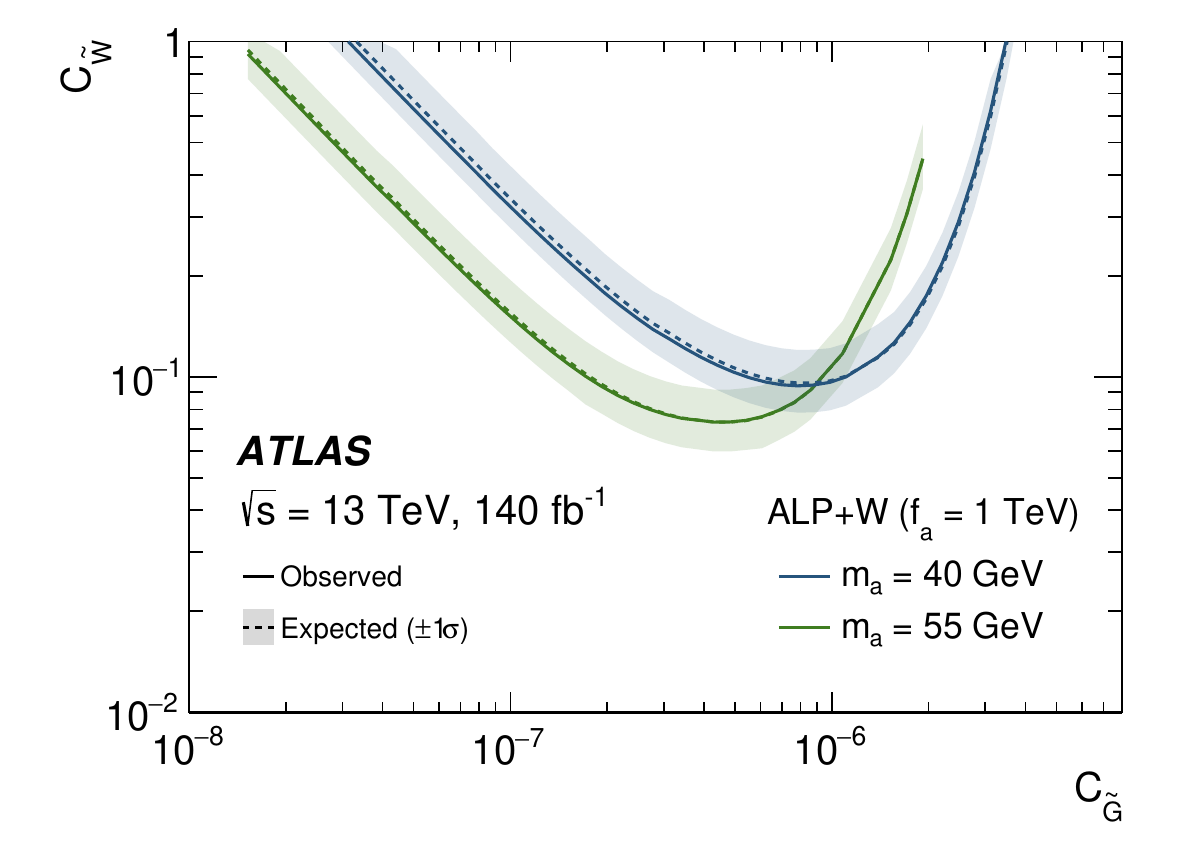}}

\end{tabular}
\caption{The expected and observed 95\% CL exclusion limits on $\sigma(qq \rightarrow Va)$ expressed in terms of the
Wilson coefficients $C_{\tilde{G}}$ and $C_{\tilde{W}}$ for the $Z$+ALP (left) and $W$+ALP (right) model considered in
this analysis, assuming a characteristic scale of $f_a = 1$~TeV~\cite{Brivio_2017}.}
\end{center}
\end{figure}

\begin{figure}[h!]
\centering
\subfloat[]{\includegraphics[width=0.5\textwidth]{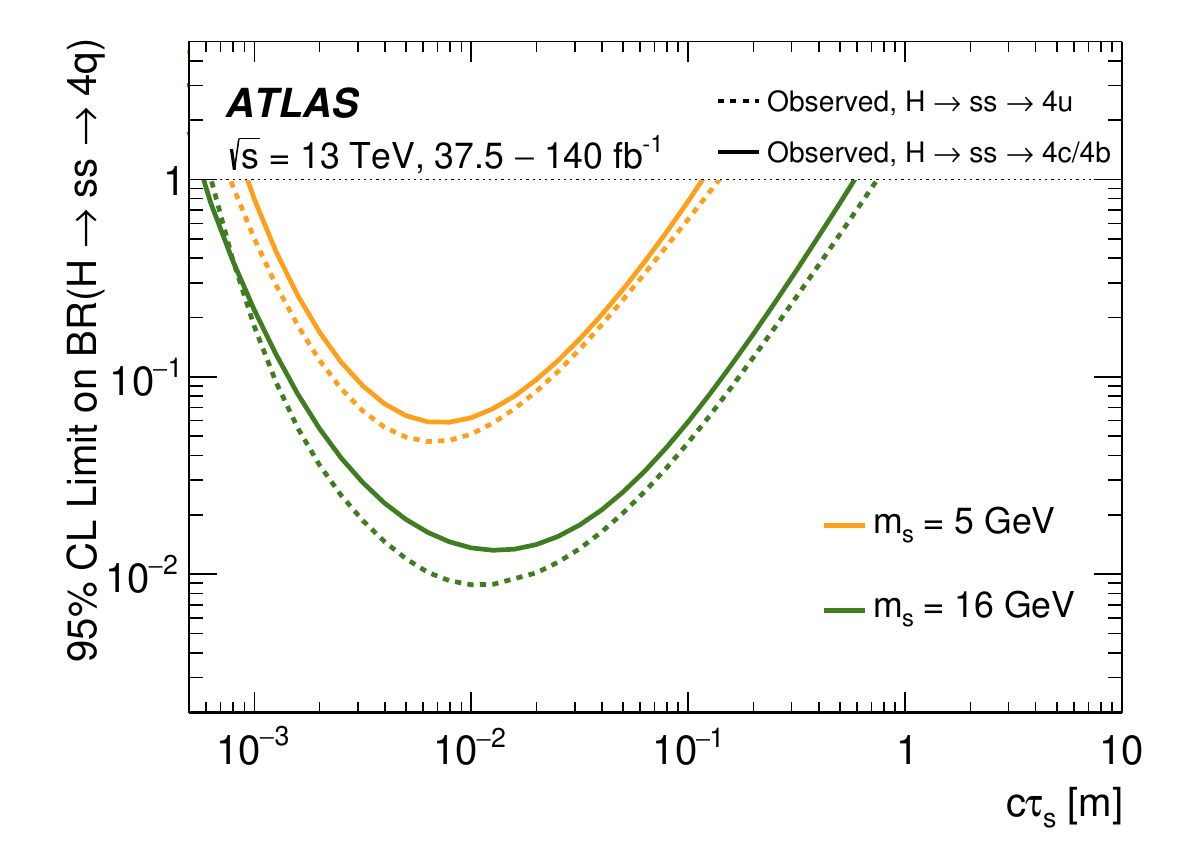} }
\subfloat[]{\includegraphics[width=0.5\textwidth]{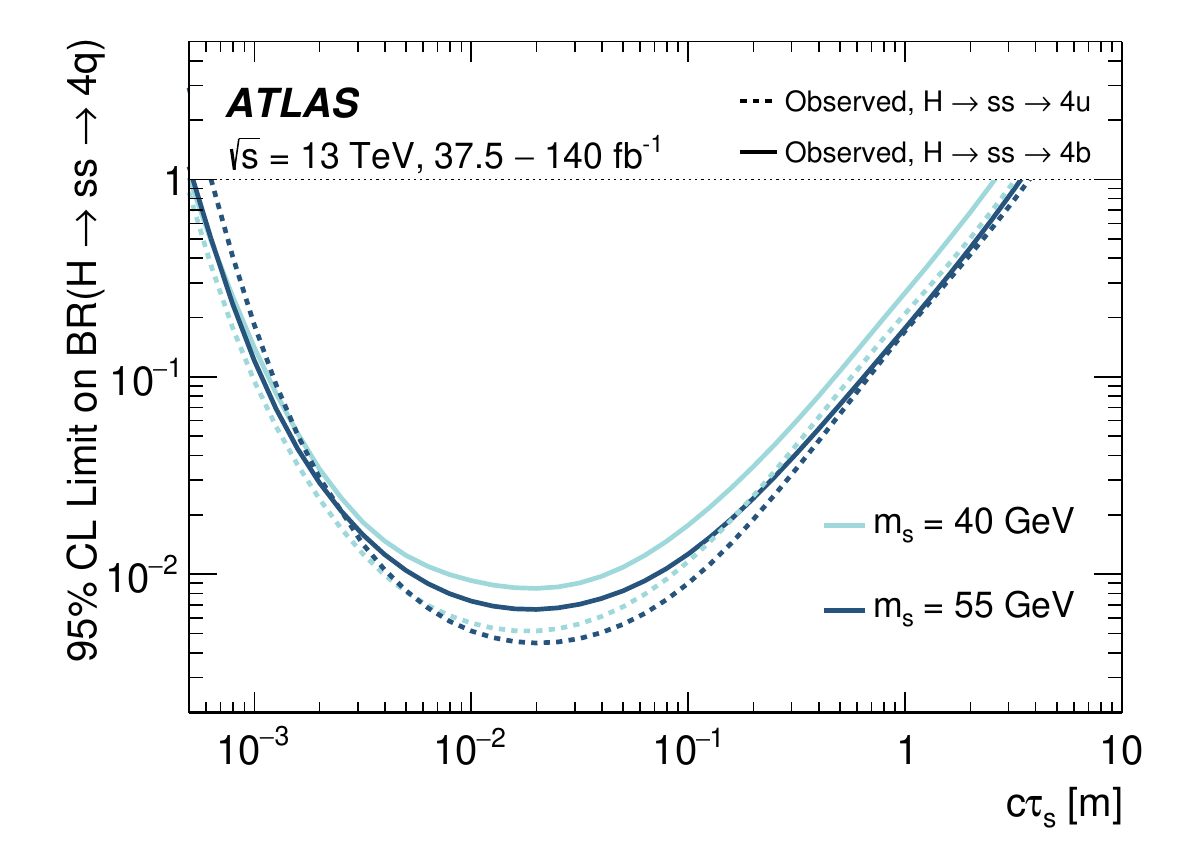} }
\caption{The 95\% confidence level limits on the Higgs boson branching ratio
shown as a function of $c\tau$, where $c$ is the speed of light and $\tau$ is
the mean proper lifetime of the long-lived particle. The limits shown are for
the Higgs boson branching ratio $H \rightarrow ss \rightarrow 4q$ for (a) light
quarks (dashed lines) with $q=u$ and heavy quarks (solid lines) with $q=c$ for $m_s=5$~GeV and $q=b$ for $m_s=16$~GeV and (b) for light quarks (dashed lines) with $q=u$ and heavy quarks (solid lines) with $q=b$ for $m_s=40,55$~GeV.
}
\label{fig:Limits_4u_vs_4b}
\end{figure}

\begin{figure}
\centering
\subfloat[]{
\includegraphics[width=0.49\textwidth]{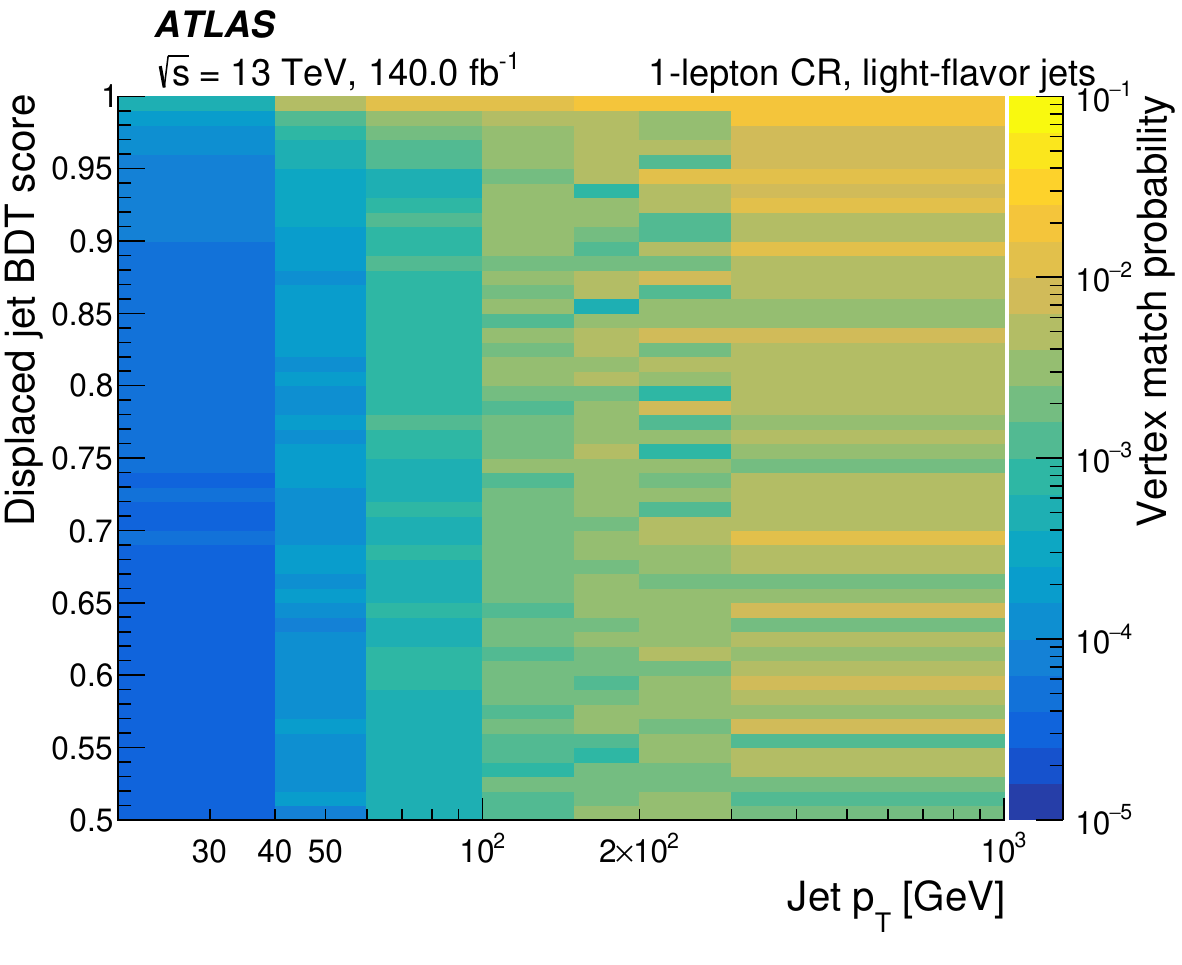}
}
\subfloat[]{
\includegraphics[width=0.49\textwidth]{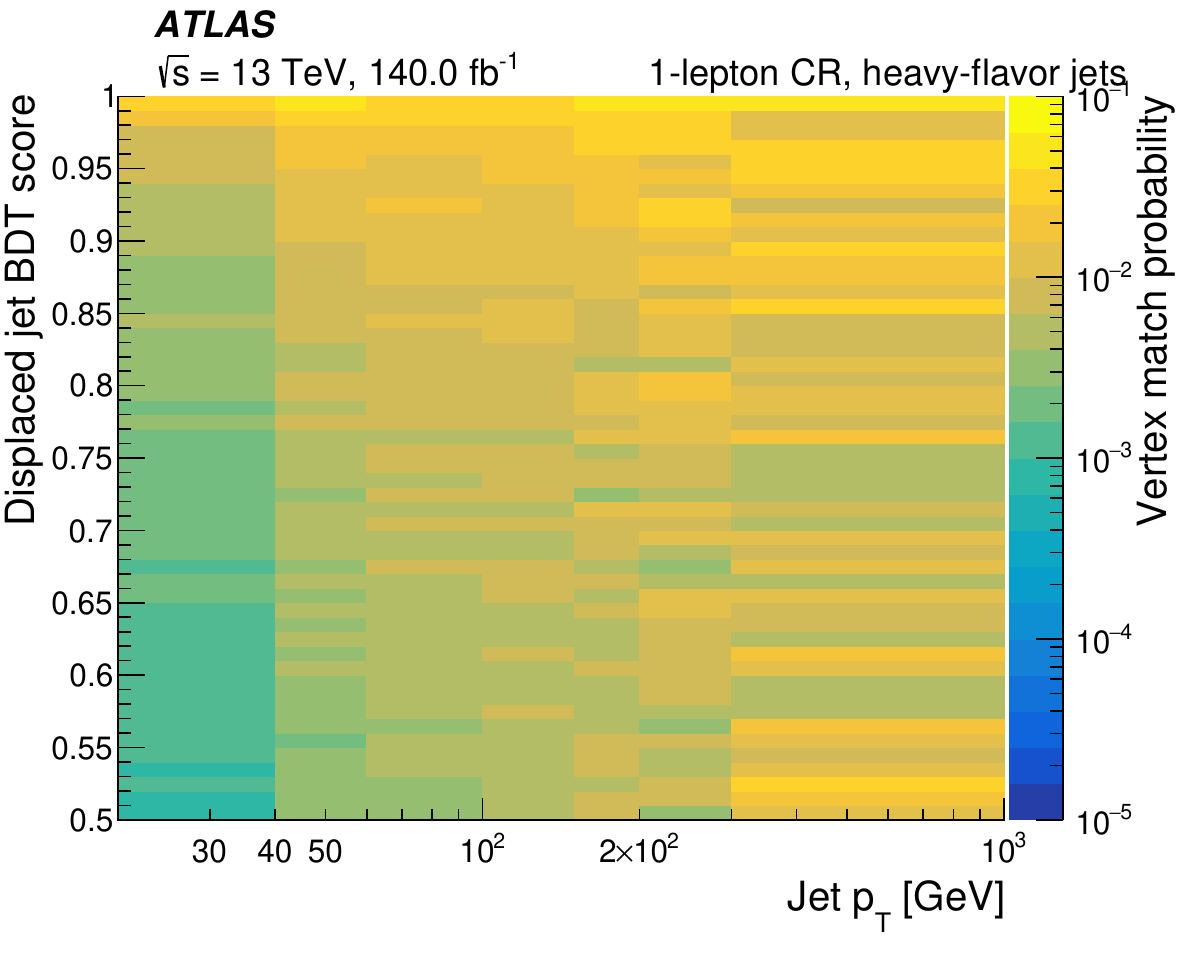}
} \\
\subfloat[]{
\includegraphics[width=0.49\textwidth]{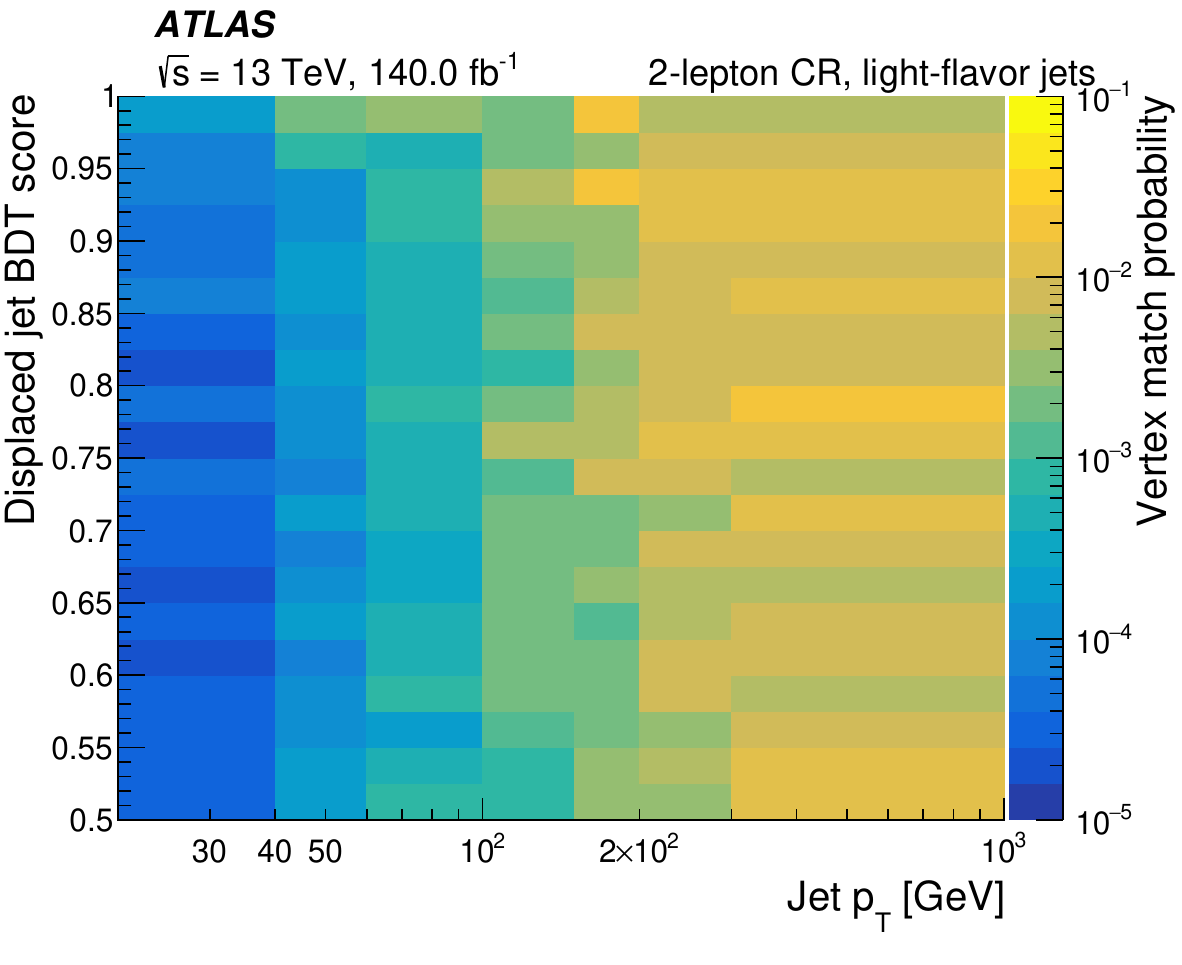}
}
\subfloat[]{
\includegraphics[width=0.49\textwidth]{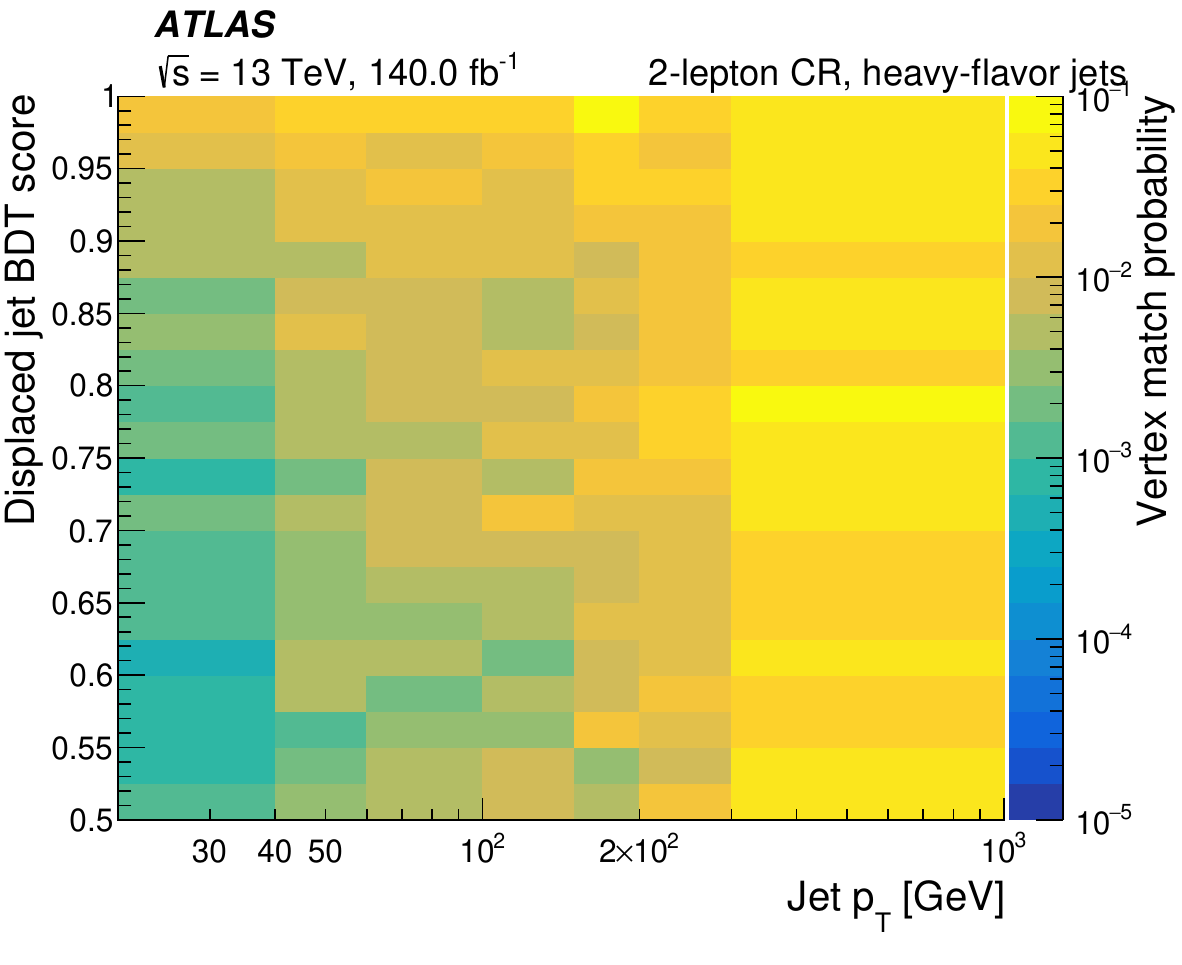}
} \\
\subfloat[]{
\includegraphics[width=0.49\textwidth]{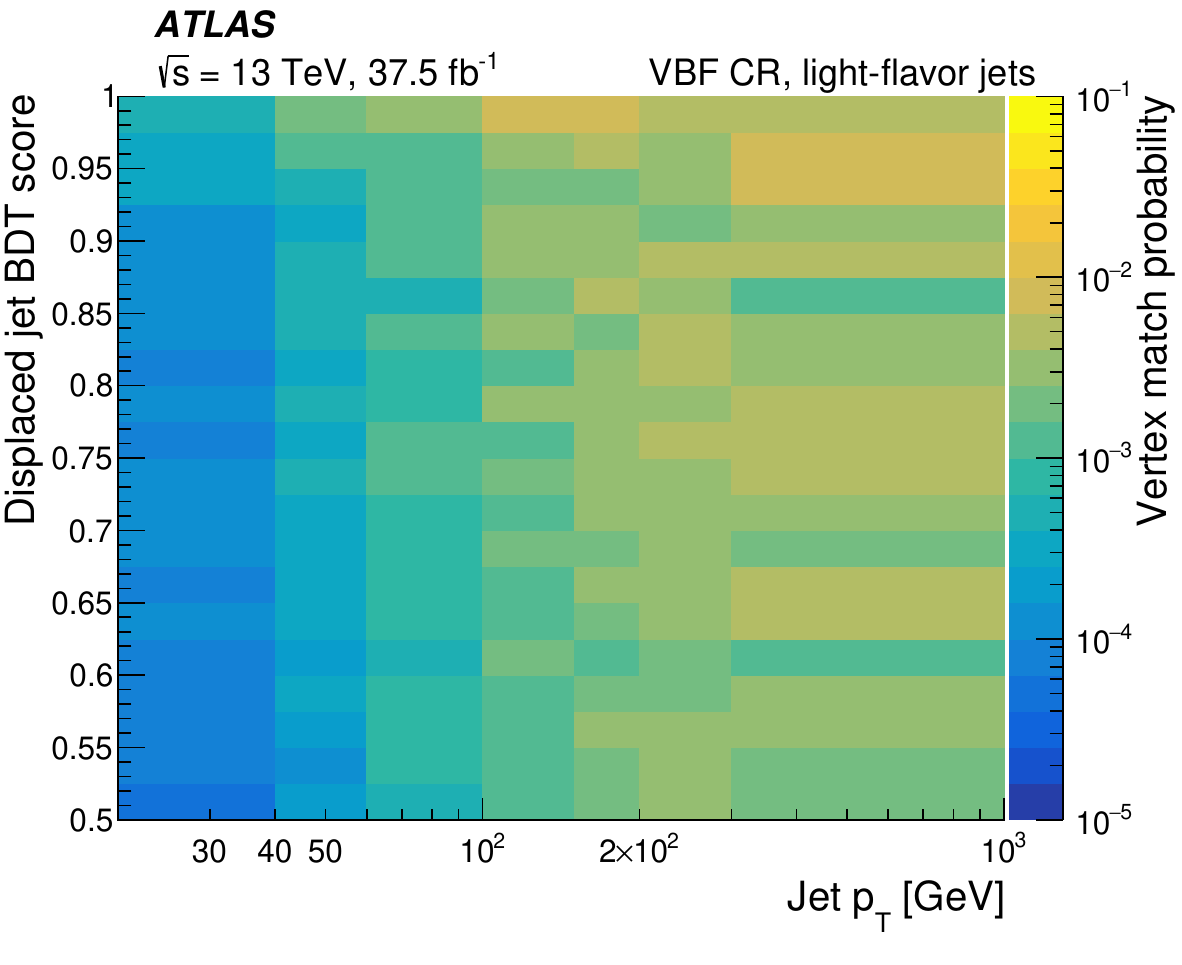}
}
\subfloat[]{
\includegraphics[width=0.49\textwidth]{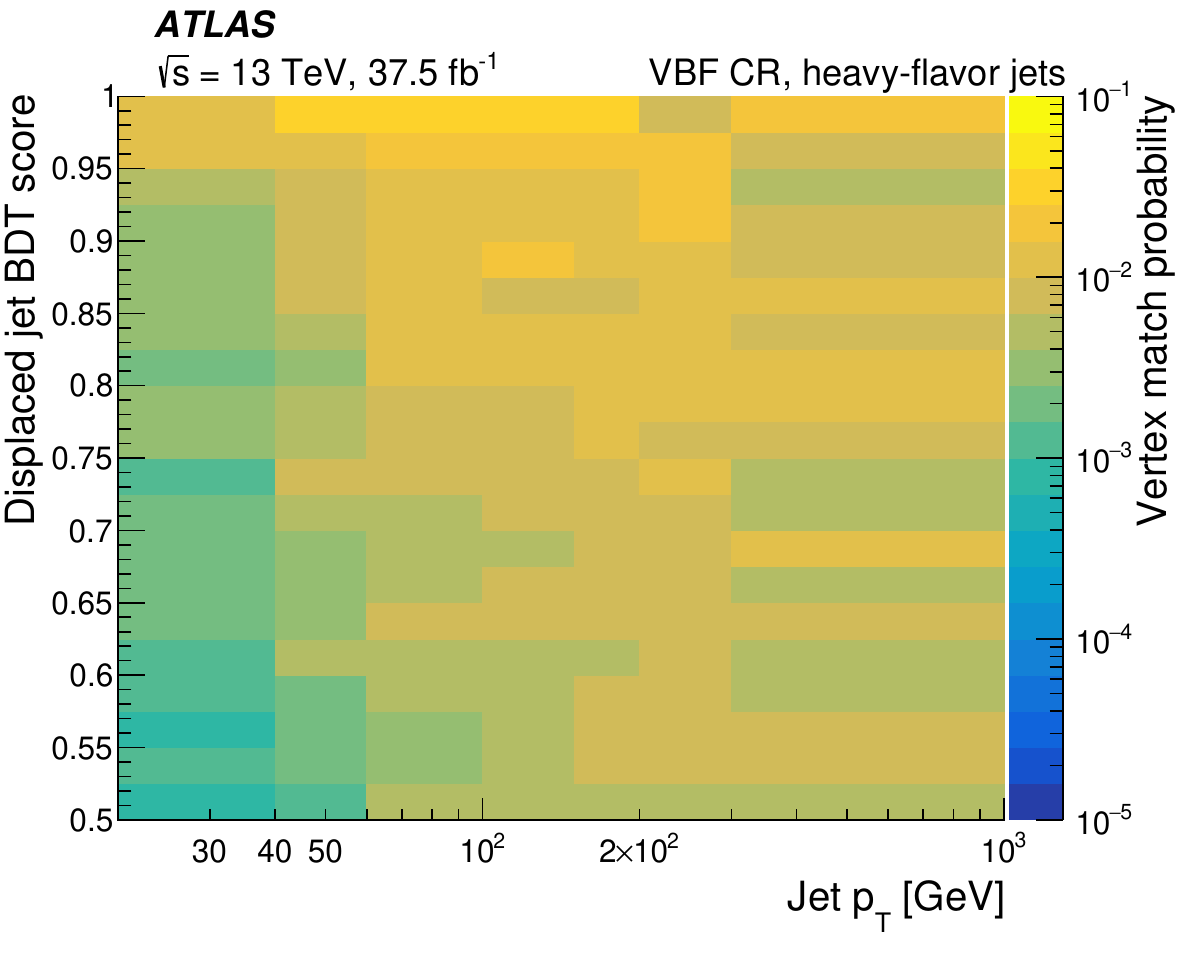}
} \\
\caption{The per-jet efficiency maps used to derive the derived using events from the (a, b) 1-lepton CR, (c, d) 2-lepton CR,
and (e, f) VBF CR. The three-dimensional maps are shown as two two-dimensional projections for
(a, c, e) light flavor jets and (b, d, f) heavy-flavor jets.}
\label{fig:efficiencyMaps}
\end{figure}

\clearpage

\clearpage
\input{atlas_authlist}
%

%
%
\clearpage

\end{document}

%% file: atlas_authlist.tex
 
\begin{flushleft}
\hypersetup{urlcolor=black}
{\Large The ATLAS Collaboration}

\bigskip

\AtlasOrcid[0000-0002-6665-4934]{G.~Aad}$^\textrm{\scriptsize 104}$,
\AtlasOrcid[0000-0001-7616-1554]{E.~Aakvaag}$^\textrm{\scriptsize 17}$,
\AtlasOrcid[0000-0002-5888-2734]{B.~Abbott}$^\textrm{\scriptsize 123}$,
\AtlasOrcid[0000-0002-0287-5869]{S.~Abdelhameed}$^\textrm{\scriptsize 119a}$,
\AtlasOrcid[0000-0002-1002-1652]{K.~Abeling}$^\textrm{\scriptsize 56}$,
\AtlasOrcid[0000-0001-5763-2760]{N.J.~Abicht}$^\textrm{\scriptsize 50}$,
\AtlasOrcid[0000-0002-8496-9294]{S.H.~Abidi}$^\textrm{\scriptsize 30}$,
\AtlasOrcid[0009-0003-6578-220X]{M.~Aboelela}$^\textrm{\scriptsize 45}$,
\AtlasOrcid[0000-0002-9987-2292]{A.~Aboulhorma}$^\textrm{\scriptsize 36e}$,
\AtlasOrcid[0000-0001-5329-6640]{H.~Abramowicz}$^\textrm{\scriptsize 154}$
% ...
\bigskip
\\

% ...
$^{17}$Department for Physics and Technology, University of Bergen, Bergen; Norway.\\
% ...
$^{30}$Physics Department, Brookhaven National Laboratory, Upton NY; United States of America.\\
% ...
$^{36}$$^{(a)}$Facult\'e des Sciences Ain Chock, R\'eseau Universitaire de Physique des Hautes Energies - Universit\'e Hassan II, Casablanca;$^{(b)}$Facult\'{e} des Sciences, Universit\'{e} Ibn-Tofail, K\'{e}nitra;$^{(c)}$Facult\'e des Sciences Semlalia, Universit\'e Cadi Ayyad, LPHEA-Marrakech;$^{(d)}$LPMR, Facult\'e des Sciences, Universit\'e Mohamed Premier, Oujda;$^{(e)}$Facult\'e des sciences, Universit\'e Mohammed V, Rabat;$^{(f)}$Institute of Applied Physics, Mohammed VI Polytechnic University, Ben Guerir; Morocco.\\
% ...
$^{45}$Physics Department, Southern Methodist University, Dallas TX; United States of America.\\
% ...
$^{50}$Fakult\"{a}t Physik , Technische Universit{\"a}t Dortmund, Dortmund; Germany.\\
% ...
$^{56}$II. Physikalisches Institut, Georg-August-Universit\"{a}t G\"ottingen, G\"ottingen; Germany.\\
% ...
$^{104}$CPPM, Aix-Marseille Universit\'e, CNRS/IN2P3, Marseille; France.\\
% ...
$^{119}$$^{(a)}$New York University Abu Dhabi, Abu Dhabi;$^{(b)}$United Arab Emirates University, Al Ain; United Arab Emirates.\\
% ...
$^{123}$Homer L. Dodge Department of Physics and Astronomy, University of Oklahoma, Norman OK; United States of America.\\
% ...
$^{154}$Raymond and Beverly Sackler School of Physics and Astronomy, Tel Aviv University, Tel Aviv; Israel.\\
% ...
$^{a}$ Also Affiliated with an institute covered by a cooperation agreement with CERN.\\
$^{b}$ Also at An-Najah National University, Nablus; Palestine.\\
$^{c}$ Also at Borough of Manhattan Community College, City University of New York, New York NY; United States of America.\\
$^{d}$ Also at Center for Interdisciplinary Research and Innovation (CIRI-AUTH), Thessaloniki; Greece.\\
$^{e}$ Also at Centro Studi e Ricerche Enrico Fermi; Italy.\\
$^{f}$ Also at CERN, Geneva; Switzerland.\\
% ...

\end{flushleft}
